\documentclass[10pt, twocolumn, a4paper]{article}

\usepackage{amsmath,amssymb}
\usepackage[a4paper, top=2.0cm, bottom=1.5cm, left=1.5cm, right=1.5cm]{geometry}
\usepackage{helvet}
\usepackage{graphicx}
\usepackage{color}
\usepackage[colorlinks=true,bookmarks=false,citecolor=blue,urlcolor=blue]{hyperref}
\usepackage{physics}

\title{Higher-order topological phase of interacting photon pairs}

\author{Andrei~A.~Stepanenko$^{1}$, Mark~D.~Lyubarov$^{1}$, Maxim~A.~Gorlach$^{1}$}

\begin{document}

\small

\twocolumn[
\begin{@twocolumnfalse}
\maketitle

{\it \small{
$^1$~School of Physics and Engineering, ITMO University, Saint Petersburg 197101, Russia
}}
\end{@twocolumnfalse}]
\ \

\textbf{Topological phases~\cite{Hasan2010,Qi2011,Ozawa2019} open a door to such intriguing phenomena as unidirectional propagation and disorder-resilient localization at a stable frequency. Recently discovered higher-order topological phases~\cite{Benalcazar2017,Schindler2018} further extend the concept of topological protection enabling versatile control over localization in multiple dimensions. Motivated by the recent advances in quantum technologies such as large coherently operating qubit ensembles~\cite{Arute2019,Wu2021}, we predict and investigate the higher-order topological phase of entangled photon pairs emerging due to the effective photon-photon interaction. Being feasible for state-of-the-art experimental capabilities, the designed model provides an interesting example of interaction-induced topological transitions in the few-particle two-dimensional system.}




Recently reported superconducting quantum processors which include 53~\cite{Arute2019} and 66 qubits~\cite{Wu2021} open a perspective of highly efficient quantum computations. However, several challenges appear on this route. One of them is unavoidable parameter spread between the fabricated qubits. A promising approach to overcome the effects of disorder is provided by the concept of topological states~\cite{Ozawa2019} which are intrinsically immune against some types of disorder being protected by the global symmetries of the structure.

While topological states of classical waves are relatively well-established~\cite{Yang2015,Huber2016,Ozawa2019}, topological protection of quantum light is less explored. At the same time, first experimental works suggest that the concepts of topology can be harnessed to protect photonic path entanglement~\cite{BlancoRedondo2018}, entanglement between photons in the spontaneous parametric four-wave mixing process~\cite{Mittal2018} or spontaneous parametric down-conversion~\cite{Wang2019a}. 

Interacting multi-photon systems are expected to bring even richer physics due to the interplay of topology and interactions. One of the platforms to investigate such physics is provided by superconducting transmon qubits~\cite{Roushan2016,Roushan2017,Cai2019,Kim2021} which exhibit a significant effective photon-photon interaction due to the anharmonicity of the qubit potential. In particular, recent experimental works have addressed the physics of bound photon pairs propagating and localizing in one-dimensional qubit arrays~\cite{Besedin2021}.

Inspired by the recent advances in quantum technologies~\cite{Arute2019,Wu2021}, we make the next conceptual step and investigate a higher-order topological phase~\cite{Benalcazar2017,Schindler2018} induced by  photon-photon interaction in a two-dimensional qubit array. Previously, higher-order topological insulators were realized experimentally for elastic waves~\cite{SerraGarcia2018}, in radiofrequencies~\cite{Imhof2018,SerraGarcia2019}, microwaves~\cite{Peterson2018} and photonics~\cite{Mittal2019a,Hassan2019}. However, higher-order topological states of quantum light have evaded their observation so far. Note that the outlined problem is qualitatively different from the recently studied physics of lattices of evanescently coupled optical waveguides with Kerr-type nonlinearity~\cite{Kirsch2021}.

\begin{figure*}[t]
    \centerline{
    \includegraphics[width = \textwidth]{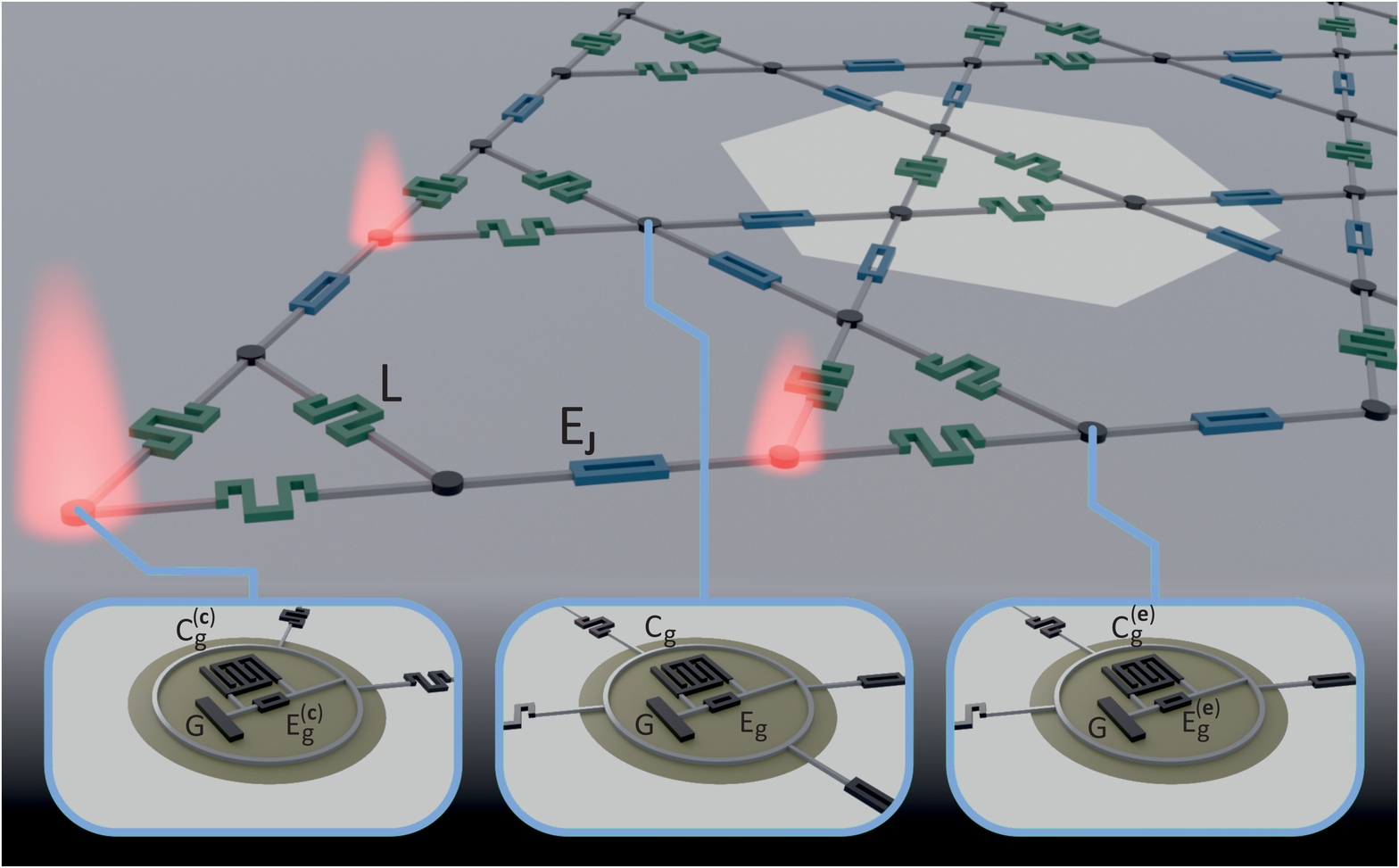}}
    \caption{\footnotesize {\bf Artistic view of the system under study.} The bulk site of the kagome lattice contains nonlinear LC oscillator made of the grounded linear capacitance $C_g$ and Josephson junction $E_g$. The sites are connected to each other either via linear inductance $L$ or via Josephson junction $E_J$. To compensate the lack of neighbors for the edge and corner sites and avoid the formation of trivial defect states, grounding elements for them are modified as shown in the insets. White hexagon shows the unit cell choice. The location of the topological corner state of bound photon pair is highlighted by red.}
    \label{fig:scheme}
\end{figure*}

In this work, we consider a two-dimensional qubit array and predict an interaction-induced topological phase transition in it. Specifically, we focus on a kagome lattice of qubits (Fig.~\ref{fig:scheme}) where each site consists of the linear capacitance  $C_g$ and Josephson junction $E_g$ which together form the nonlinear LC oscillator. The qubits in turn are connected to each other either via linear inductances $L$ or via nonlinear couplings formed by the Josephson junctions $E_J$. To compensate the lack of neighbors for the edge and corner sites, we adjust the magnitude of the grounding elements for them accordingly. 

To describe multi-photon excitations in the proposed structure, we introduce flux variables $\phi_{m,n}^{\alpha}$, where $\alpha = a,b,c$ denotes one of the three sublattices, and $(m,n)$ indices enumerate unit cells. The quantization of our model is outlined in Supplementary Note~1. The derived Hamiltonian corresponds to the extended Bose-Hubbard model:
\begin{eqnarray}
\hat{H} &=& f_0\sum_{m,n,\alpha} \hat{n}_{m,n}^{\alpha}  + \hat{V}_{J} +\hat{V}_{I}\:, \label{H in text}\\
\hat{V}_{J} &=& \sum_{m,n,\alpha\neq\beta}\left[J_L(\hat{a}_{m,n}^{\alpha})^\dagger \hat{a}_{m,n}^{\beta}
+J_J\sum_{m',n'}(\hat{a}_{m,n}^{\alpha})^\dagger \hat{a}_{m',n'}^{\beta}\right]\\
\hat{V}_{I} &=&  \frac{E^k}{2} \sum_{m,n,\alpha} \hat{n}_{m,n}^{\alpha}(\hat{n}_{m,n}^{\alpha}-1)\nonumber\\
&&
+\frac{T}{2}\sum_{m,n,m',n',\alpha\neq\beta}(\hat{a}_{m,n}^{\alpha}\hat{a}_{m,n}^{\alpha})^\dagger \hat{a}_{m',n'}^{\beta}\hat{a}_{m',n'}^{\beta}
\nonumber\\
&&+\frac{J^D}{\sqrt{2}}\sum_{m,n,m',n',\alpha\neq\beta}(\hat{a}_{m,n}^{\alpha})^\dagger(\hat{n}_{m,n}^{\alpha}+\hat{n}_{m',n'}^{\beta})\hat{a}_{m',n'}^{\beta}
\nonumber\\
&&+E^{ck}\sum_{m,n,m',n',\alpha\neq\beta}\hat{n}_{m,n}^{\alpha}\hat{n}_{m',n'}^{\beta}\:,
\label{eq:bh_Hamiltonian}
\end{eqnarray}
where $\hat{a}_{m,n}^{\alpha}$ is an annihilation operator at site $\alpha$ of $(m,n)$ unit cell,  $\hat{n}_{m,n}^{\alpha} = (\hat{a}_{m,n}^{\alpha})^\dagger\hat{a}_{m,n}^{\alpha}$, and $(m',n',\beta)$ indices correspond to the site adjacent to  $(m,n,\alpha)$. Here, parameters $f_0$, $J_L=-Z/2L$, $J_J=-ZE_J/2+E_JZ^2/4$ and $E^k$ correspond to the eigenfrequency, intra- and inter-cell couplings and on-site anharmonicity, respectively, which appear in the conventional Bose-Hubbard model. Note that all parameters are renormalized to have the dimensionality of frequency via the substitution $C\phi_0^2/h \rightarrow C$, $Lh/\phi_0^2 \rightarrow L$, $Zh/\phi_0^2 \rightarrow Z$, $E_{J,g}/h \rightarrow E_{J,g}$, where $\phi_0=h/(2e)$ is the reduced flux quantum and $e$ is the elementary charge (see further details in the Methods section). The additional terms $T$, $J^D$ and $E^{ck}$ are responsible for the direct two-photon tunneling, density-dependent coupling and cross-Kerr interaction which have been previously investigated in the context of one-dimensional problems in Refs.~\cite{Stepanenko2020,Stepanenko-PRAppl,Gorlach2017}.



To grasp the physics of the designed higher-order topological insulator in the two-photon regime, we consider the limit of strong anharmonicity $|E^k|\gg {\rm max}\left(J_L,J_J\right)$. In that case, bound two-photon states are well-separated spectrally from the rest of the two-photon states arising regardless of the sign of $E^k$~\cite{Winkler2006}.


The topological properties of the system rely on the behavior of the bulk modes found from the Schr\"odinger equation $\hat{H}|\psi\rangle = \varepsilon|\psi\rangle $, which is solved numerically for a finite system. 

\begin{figure}[ht]
    \centerline{
    \includegraphics[width = 0.49\textwidth]{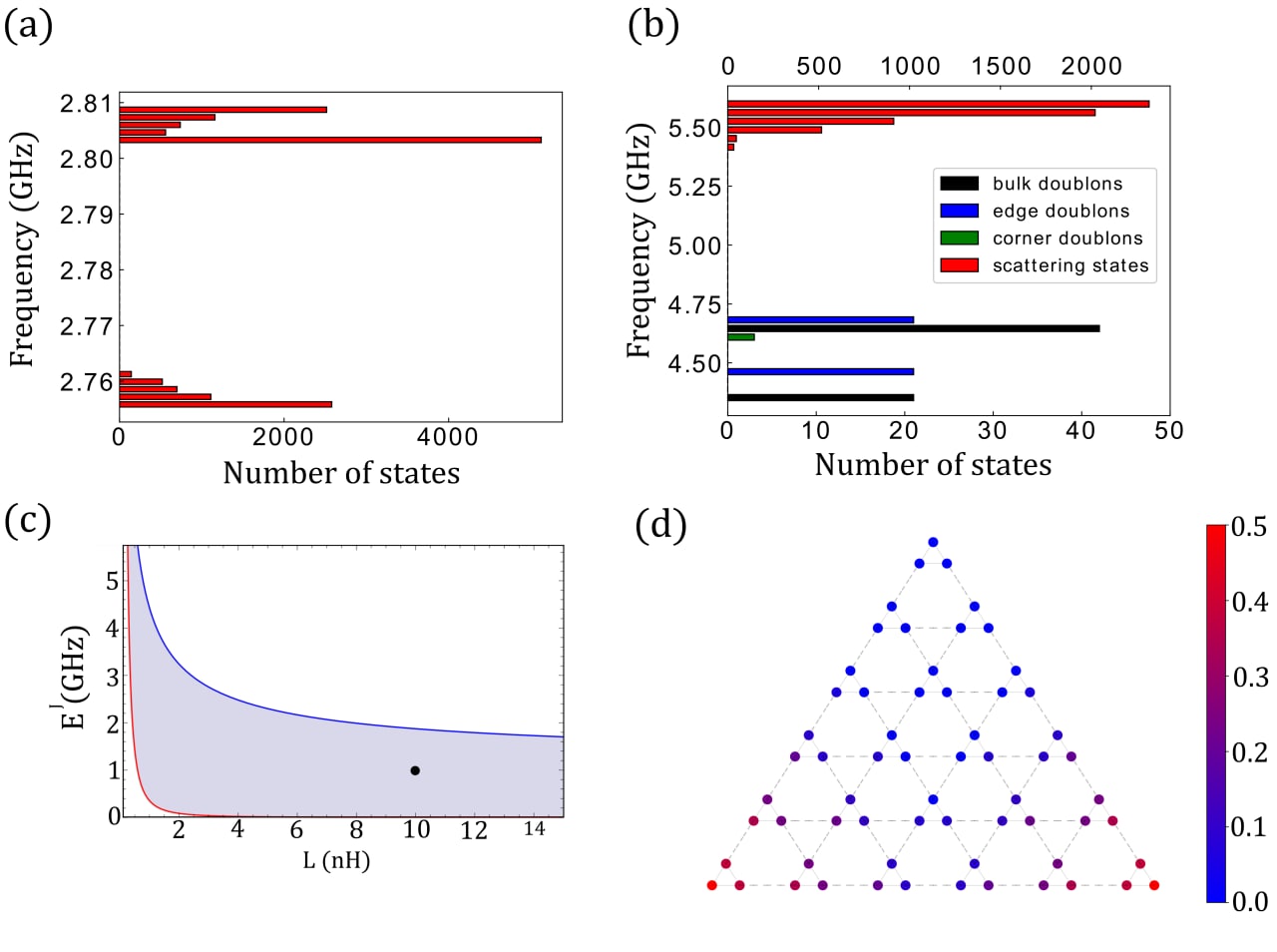}}
    \caption{\footnotesize {\bf Single- and two-photon spectra.} Density of states calculated for the qubit array for the following parameters: $L = 10$~nH, $E_J = 1$~GHz, $C = 0.02$~fF, $E_g = 10$~GHz for ({\bf a}) single-photon case and the system of $n=120$ qubits, where $n$ is the number of unit cells along one side of the triangle, and ({\bf b}) two-photon case and the system size $n=8$. Bulk doublon states are accompanied by the doublon edge and corner states. The scales on the top and on the bottom of the plot correspond  to the scattering states and doublon states, respectively. In both panels ({\bf a}), ({\bf b}) the grounding elements $E_g, C_g$ of the edge and corner sites are modified to avoid the formation of trivial Tamm-like states. ({\bf c}), Phase diagram of topological regimes calculated for $C = 0.02$~fF, $E_g = 10$~GHz. Gray filling shows the regime when the single-photon model is trivial, while the two-photon system is topological. Black dot indicates the normalized parameters chosen for the calculations below:  $L = 10$~nH, $E_J = 1$~GHz, $C = 0.02$~fF, $E_g = 10$~GHz. ({\bf d}), Topological doublon corner state for the system of size $n=6$. The colorbar shows the value  $S_n = \text{max}(0, 1+\ln(<n_i>/2)/\sigma)$, where $\sigma=35$ is an auxiliary parameter chosen to highlight the exponential decay of the corner state.}
    \label{fig:1-2_phot_comparison}
\end{figure}

First, we revisit a single-photon case that is analogous to the well-studied classical scenario~\cite{Ezawa2018,Xue2018,Ni2018}. Overall, the spectrum consists of bulk, edge and corner states, Fig.~\ref{fig:1-2_phot_comparison}{\bf a}. The latter two types of states appear only in the topological phase which is controlled by the ratio of intra- and intercell coupling constants: $J_L/J_J<1$. The ratio $J_L/J_J=1$ corresponds to the topological transition point accompanied by closing and reopening of a complete bandgap. This condition translates into the following relation between the parameters of the system:
\begin{eqnarray}
|LE_J(Z-2)/2|=1\:,
\label{eq:sp_tt}
\end{eqnarray}
that defines a curve on $L$-$E_J$ plane shown by blue in  Fig.~\ref{fig:1-2_phot_comparison}{\bf c}. The region below this line corresponds to the topologically trivial situation when only bulk single-photon states are present [Fig.~\ref{fig:1-2_phot_comparison}{\bf a}].

At the same time, the two-photon spectrum (Fig. \ref{fig:1-2_phot_comparison}{\bf b}) is much richer containing scattering states, single-photon edge states with one of the photons localized at the edge, as well as bound photon pairs (doublons). In turn, the doublon states feature the same hierarchy exhibiting bulk, edge, and corner states. The scattering states and single-photon edge modes inherit their topological properties from the single-particle case, being topologically trivial. However, the situation is different for the case of doublons.

To demonstrate this, we plot the condition of topological transition for the doublon bands in Fig.~\ref{fig:1-2_phot_comparison}{\bf c} by the red curve (see Methods for the details of derivation). The domain above this curve corresponds to the topological regime. Hence, the area between the two curves shaded in Fig.~\ref{fig:1-2_phot_comparison}{\bf c} by gray realizes the situation when the single-photon bands are trivial, while bound two-photon states are topological. In the other words, this signals the topological transition happening in the system when the second photon is introduced.

\begin{figure}[ht]
    \centerline{
    \includegraphics[width = 0.49\textwidth]{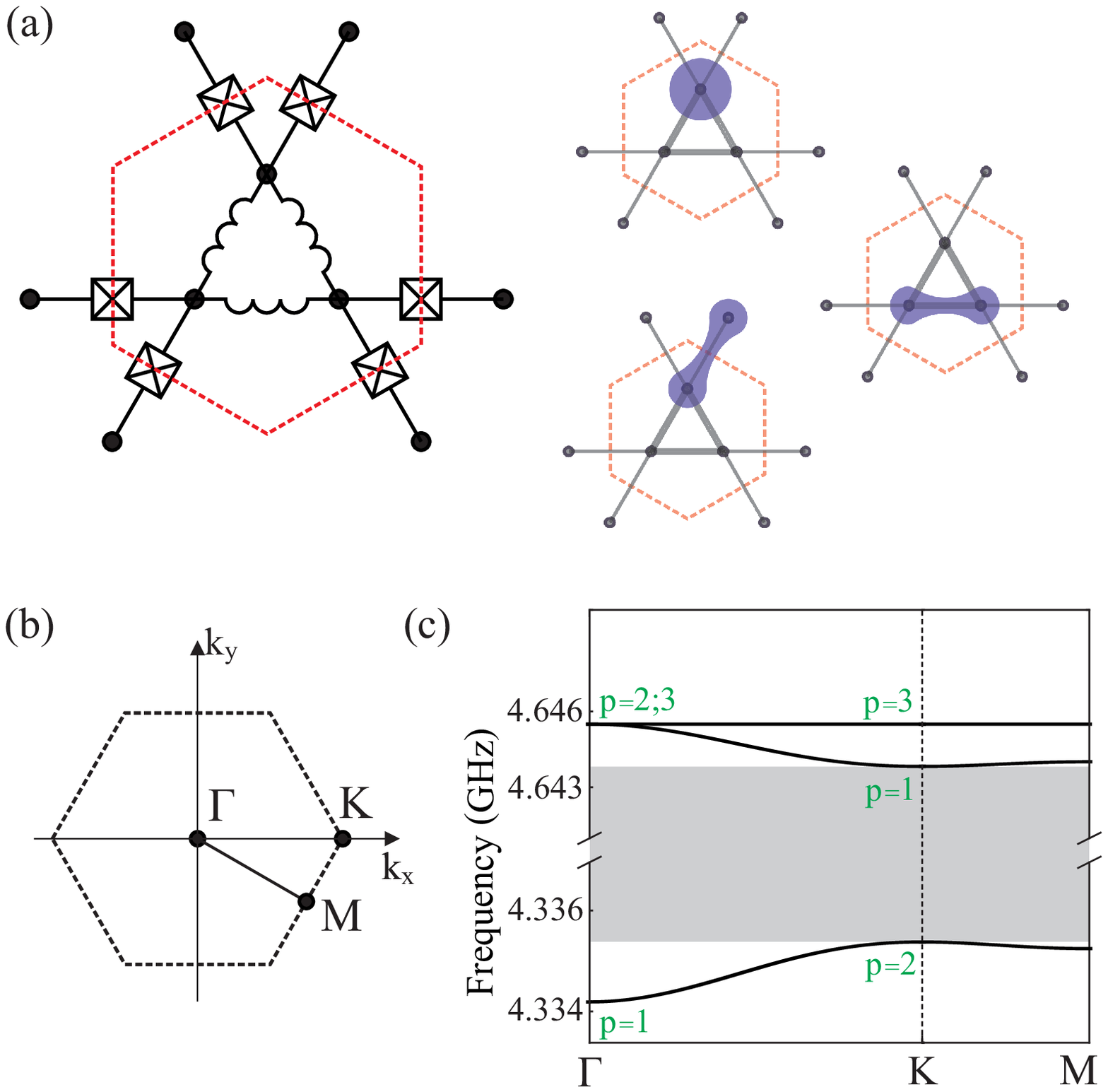}}
    \caption{\footnotesize {\bf Topology of doublon bands.}  ({\bf a}), The unit cell contains tree grounded qubits ($E_g, C_g$) connected through inductance $L$. The unit cells are connected via Josephson junctions $E_J$. Inset shows the examples of doublon state and two-photon states with photons localized in the neighbouring qubits. ({\bf b}), The first Brillouin zone for the designed lattice. ({\bf c}), The band structure of the system with symmetry indicators $p$ calculated in the high-symmetry points of the first Brillouin zone. The gray filling shows a complete photonic bandgap.}
    \label{fig:topology_doublons}
\end{figure}

To assess the properties of the two-photon states, we employ the second-order perturbation theory dealing with the truncated basis that involves two-photon states with the photons localized in the same site or in the neighboring ones (right part of the Fig.~\ref{fig:topology_doublons}{\bf a}). This approach yields doublon effective hopping amplitudes which include the contributions from the conventional linear tunneling as well as higher-order processes. Further details of our analysis are provided in Methods.


To explicitly check trivial and topological origin of single- and two-photon states, respectively, we consider a periodic system with the unit cell and the Brillouin zone shown in Fig.~\ref{fig:topology_doublons}{\bf a,b}, respectively. Extracting the periodic part of the Bloch function, we examine its behavior under $2\pi/3$ rotations described by the operator 
\begin{eqnarray}
\hat{R}_3 &=& \left(
\begin{array}{ccc}
     0&0&1  \\
     1&0&0  \\
     0&1&0  
\end{array}
\right)\:.
\end{eqnarray}
Since $\hat{R}_3$ matrix commutes with the Hamiltonian in $K$ point, all eigenstates in this point can be labelled by $\hat{R}_3$ eigenvalues which can be presented in the form $\exp[2\pi i(p-1)/3]$ where $p = 1,2,3$. The calculated symmetry indices are indicated for each of the doublon bands in $\Gamma$ and $K$ points in Fig.~\ref{fig:topology_doublons}{\bf c}.

Given the set of symmetry indices, we construct the topological invariant~\cite{Benalcazar2019} $\chi = (\#K_1-\#\Gamma_1,\#K_2-\#\Gamma_2)$, where $\#\Pi_p$ is a number of the eigenmodes with a given symmetry index $p$ below the bandgap under consideration in the point $\Pi$ of the Brillouin zone.

Applying this approach to the single-photon bands, as further detailed in Supplementary Note 2, we recover the single-photon topological invariant $\chi = (0,0)$, which indicates the absence of topological corner states aligning with our previous estimation based on the ratio of the coupling amplitudes.

In contrast, we observe the band inversion in the two-photon case [Fig.~\ref{fig:topology_doublons}{\bf c}], obtaining nonzero topological invariant  $\chi = (-1,1)$ signaling nonzero corner charge and topological corner state. Simulating the finite array, we indeed recover such a state [Fig.~\ref{fig:1-2_phot_comparison}{\bf d}]. To visualize the complicated structure of the wave function, we plot the expectation value of the photon number $<n_i>$ for each of the lattice sites. The calculated profile clearly shows a significant decay proving the formation of the localized state. Due to its two-photon nature, the corner state also features nontrivial entanglement properties which are not captured by the chosen visualization scheme.

It should be stressed that the predicted two-photon topological corner state should be distinguished from the interaction-induced trivial Tamm states formed due to the nonlinear detuning of the edge and corner sites arising because of the lack of neighbors. To eliminate such states, we introduce additional modification of the boundary elements $L_g^{(e,c)}$ and $E_g^{(e,c)}$.

To summarise, we have demonstrated a second-order topological phase induced by the effective photon-photon interactions in the qubit network. On-site qubit anharmonicity allows us to achieve the formation of bound photon pairs while the nonlinear couplings between the sites via Josephson junctions enable topological transition. Our results reveal the role of interactions in the formation of multi-particle topological phases protected by $C_3$ lattice symmetry. Finally, the proposed design is feasible for the state-of-the-art technologies involving the arrays of superconducting qubits and hence opens new horizons in the development of topologically protected quantum chips and transmission lines.





{\scriptsize

\bibliographystyle{naturemag}
\bibliography{lib}

}

\newpage
\section*{\small Methods}
{\scriptsize

\textbf{Parameters of the Bose-Hubbard Hamiltonian.}
The doubled eigenfrequency $f_0 = (C_g Z)^{-1}-Z^2(E_g+4E_j)/8$ together with anharmonicity $E^k= -Z^2(E_g + 2 E_j)/8$ define the reference two-photon level. The intra-cell coupling is described by the amplitude $J_L =-Z/2L$. Due to the presence of a Josephson junction, the inter-cell coupling contains single-photon hopping $J_J= -ZE_J/2+E_JZ^2/4$ as well as a direct two-photon hopping $T = -Z^2E_J/8$, density-dependent coupling $J^D = Z^2E_J/\sqrt{32}$, and cross-Kerr interaction $E^{ck} = -Z^2E_J/4$, where $Z = (C_g (E_g + 2 (1/L + E_J)))^{-1/2}$ is the impedance.
\newline

\textbf{Density of states.}
The spectra presented in Fig. \ref{fig:1-2_phot_comparison} a, b were calculated numerically by diagonalizing the Hamiltonian Eq.~\eqref{H in text} for the system with $1$ or $2$ photons and the sizes $n=120$ and $n=8$, respectively ($n$ is the number of qubits along the edge of the triangle, the system of size $n$ consists of $N=3n(n+1)/2$ qubits in total). In the single-photon case, the Hilbert space includes $N$ different single-photon states. Thus, the Hamiltonian matrix $H^1_{ij}=\langle i|\hat{H}|j\rangle$, $i,j=1\dots N$ has the size $N\times N$, and its eigenvalues are found by the exact diagonalization. The spectrum presented in Fig.~\ref{fig:1-2_phot_comparison}{\bf a} provides the number of eigenstates in 35 identical frequency intervals.

In the two-photon case the procedure is analogous, but now the basis in the Hilbert space consists of $N^2$ states, since both photons can occupy one of $N$ qubits. Thus, we construct the $N^2\times N^2$ Hamiltonian matrix and obtain $N^2$ eigenvalues after diagonalization. The resulting spectrum consists of both fermionic and bosonic states, so we have to sort these states according to the symmetry with respect to the permutation, to finally get $N(N+1)/2$ two-photon eigenstates. The spectrum presented in Fig. \ref{fig:1-2_phot_comparison} b is these eigenstates divided by $40$ equidistant frequency bins.
\newline

\textbf{Reduced Bloch Hamiltonian}
To analyze topological behaviour of bound photon pairs, we consider the states localized at the same site $|2_{m,n}^{a}\rangle$,$|2_{m,n}^{b}\rangle$,$|2_{m,n}^{c}\rangle$ and states with photons in the neighbouring sites: either from the same unit cell $|1_{m,n}^{a}1_{m,n}^{b}\rangle; |1_{m,n}^{a}1_{m,n}^{c}\rangle; |1_{m,n}^{b}1_{m,n}^{c}\rangle$ or from the neighboring unit cells $|1_{m,n}^{a}1_{m-1,n}^{b}\rangle; |1_{m,n}^{a}1_{m-1,n-1}^{c}\rangle; |1_{m,n}^{b}1_{m,n-1}^{c}\rangle$.
Taking into account the translation symmetry of the lattice, we introduce doublon Bloch wave vector $|\psi_{m,n+1}\rangle = \exp(ik_x)|\psi_{m,n}\rangle$ and $|\psi_{m+1,n}\rangle = \exp(ik_x/2+i\sqrt{3}k_y/2)|\psi_{m,n}\rangle$ and obtain the matrix of Bloch Hamiltonian
\begin{eqnarray}
\left(
\begin{array}{ccccccccc}
    H_{0}^{k} & T_1 & T_2 & 2J_L & 2J_L & 0 & 2J_0 & 2J_0 & 0 \\
    T_1^* & H_{0}^{k} & T_3 & 2J_L & 0 & 2J_L & 2J_1 & 0 & 2J_0 \\
    T_2^* & T_3^* & H_{0}^{k} & 0 & 2J_L & 2J_L & 0 & 2J_2 & 2J_3 \\
    J_L & J_L & 0 & H_{0} & J_L  & J_L & 0 & 0 & 0 \\
    J_L & 0 & J_L & J_L & H_{0} & J_L & 0 & 0 & 0 \\
    0 & J_L & J_L& J_L & J_L & H_{0} & 0 & 0 & 0 \\
    J_0 & J_1^* & 0 & 0 & 0 & 0 & H_{0}^{ck} & J_J & J_5   \\
    J_0 & 0 & J_2^* & 0 & 0 & 0 & J_J & H_{0}^{ck} & J_5   \\
    0 & J_0 & J_3^* & 0 & 0 & 0 & J_5^* & J_5^* & H_{0}^{ck} \\
\end{array}
\right)\:,
\label{eq:bloch_hamiltonian}
\end{eqnarray}
where $T_1 = Te^{i\phi}, T_2 = Te^{i(\phi-\theta)}, T_3 = Te^{-i\theta} $ are direct two-photon couplings and tunneling amplitudes $J_0 = J_J+J^D, J_1 = J_0e^{-i\phi}, J_2 = J_0e^{i(\theta-\phi)}, J_3 = J_0e^{i\theta}$, $J_5=J_Je^{i\phi}$ describe single-photon hopping.
Here, $\phi = (k_x+\sqrt{3}k_y)/2$ and $\theta = k_x$ are phases, and $H_{0} = 2(\omega+\delta\omega)$, $H_{0}^{k} =2(\omega+\delta\omega)+E^k$, $H_{0}^{ck} = 2(\omega+\delta\omega)+E^{ck}$ are energies.
\newline

\textbf{Perturbation theory.}
Now, we derive the effective Hamiltonian employing the degenerate second-order perturbation theory\cite{Bir}. The result reads:
\begin{eqnarray}
H^{(eff)} = H^{(0)} + H^{(1)} + H^{(2)}\:.
\end{eqnarray}
Here, the unperturbed part describes isolated qubits:
\begin{eqnarray}
H^{(0)}=\left(
\begin{array}{ccc}
    H_{0}^k & 0 & 0 \\
    0 & H_{0}^k & 0\\
    0 & 0 & H_{0}^k 
\end{array}
\right)\:,
\end{eqnarray}
the first-order perturbation
\begin{eqnarray}
H^{(1)}=\left(
\begin{array}{ccc}
    0 & Te^{i\phi} & Te^{i(\phi-\theta)} \\
    Te^{-i\phi} & 0 & Te^{-i\theta} \\
    Te^{i(\theta-\phi)} & Te^{i\theta}&  0
\end{array}
\right)\:,
\end{eqnarray}
and the second-order correction
\begin{eqnarray}
H^{(2)}=\left(
\begin{array}{ccc}
    2(H_2^a+H_2^b) & H_2^a+H_2^b e^{i\phi} & H_2^a+H_2^be^{i(\phi-\theta)} \\
    H_2^a+H_2^be^{-i\phi} & 2(H_2^a+H_2^b) & H_2^a+H_2^be^{-i\theta}\\
    H_2^a+H_2^be^{i(\theta-\phi)}& H_2^a+H_2^be^{i\theta} &  2(H_2^a+H_2^b)
\end{array}
\right)\:,
\end{eqnarray}
where
\begin{eqnarray}
H_2^a &=& 2\frac{J_L^2}{H_{0}^k-H_{0}}\:,\\
H_2^b &=& 2\frac{J_{0}^2}{H_{0}^k-H_{0}^{ck}}\:.
\end{eqnarray}

The effective couplings $J_1^{(eff)} = H_2^a$ and $J_2^{(eff)} = H_2^b+T$. Similarly to the single-particle case, the topological transition occurs under the condition $|J_1^{(eff)}/J_2^{(eff)}| = 1$
which is shown by the red curve in Fig. \ref{fig:1-2_phot_comparison}c.
\newline

\textbf{Doublon dispersion}
We perform a calculation of the dispersion of the doublons based on the exact diagonalization of the reduced Bloch Hamiltonian Eq.~\ref{eq:bloch_hamiltonian}. To this end, we take a uniform grid of points in reciprocal space and solve eigenvalue problem for  each of them. Then focusing on doublon states, we plot the diagram Fig.~\ref{fig:topology_doublons}c.

}

\subsection*{\small Acknowledgments}
The authors acknowledge Dmitrii Stepanenko for the help in illustration preparation.
This work was supported by the Russian Science Foundation (grant No.~21-72-10107).
A.A.S. acknowledges partial support by the Foundation for the Advancement of Theoretical Physics and Mathematics ``Basis", Government of the Russian Federation (Grant No.~SP-3707.2021.5) and Gennady Komissarov Foundation.

\subsection*{\small Author contributions}
M.G. supervised the project. A.S. and M.L. developed the theoretical models and performed numerical studies. A.S. and M.G. prepared the paper with input from M.L.

\subsection*{\small Data availability}
The data that support the findings of this study are available from the corresponding author upon request.

\subsection*{\small Competing interests}
The authors declare that they have no competing interests.

\subsection*{\small Additional information}
Correspondence and requests for materials should be addressed to M.G. (email: m.gorlach@metalab.ifmo.ru) or A.S. (email: andrey.stepanenko@metalab.ifmo.ru)

\newpage
\onecolumn
{\centering
\section*{Supplementary Materials}}

\section{Derivation of the circuit Hamiltonian}
\begin{figure*}[ht]
    \centerline{
    \includegraphics[width = 0.2\textwidth]{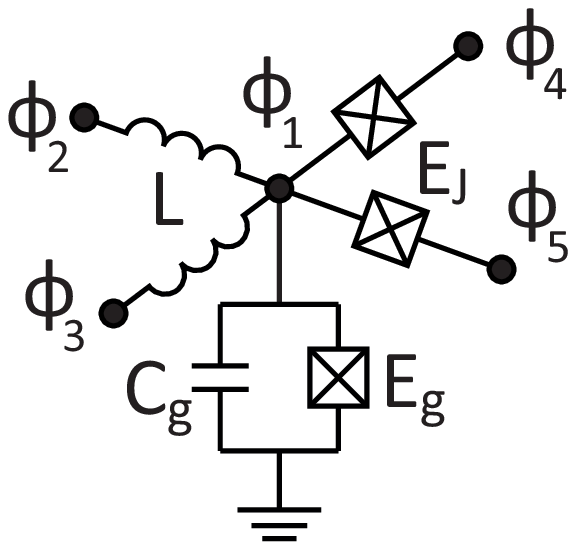}}
    \caption{\footnotesize {\bf The scheme of the node of the qubit network.} The nonlinear LC oscillator contains the grounded linear capacitance $C_g$ and Josephson junction $E_g$. $\phi_i$ is a flux in the node $i$. The intra-cell coupling (the linear inductance $L$) connects node 1 with 2 or 3 while inter-cell connection (Josephson junction $E_J$) couples the node 1 to 4 or 5.}
    \label{fig:2D_scheme}
\end{figure*}

In our analysis of the circuit equations we distinguish three situations: bulk, edge and corner nodes.

\subsection{Bulk node}
In this Note, we derive the extended Bose-Hubbard Hamiltonian for the array of qubits. We consider one bulk site of the qubit network shown in Fig.~\ref{fig:2D_scheme} described by the Lagrangian 
\begin{equation}
    \mathcal{L}=\frac{C_g}{2}\dot{\phi}_1^2 + E_g \cos{\phi_1} - \frac{(\phi_1-\phi_2)^2}{2L}-\frac{(\phi_1-\phi_3)^2}{2L}+E_J\cos{(\phi_1-\phi_4)}+E_J\cos{(\phi_1-\phi_5)}\:,
    \label{Lagrangian}
\end{equation}
where node flux variable $\phi_i^\alpha\equiv\int^tV_i^\alpha(t')dt'$ and $V_i^\alpha(t')$ is the voltage in this node with respect to the ground. Here $L$ is the inductance between nodes 1 and 2 or 3, and $E_J$ is the Josephson energy for the junction between the nodes 1 and 4 or 5. $C_g$ and $E_g$ are the capacitance and the energy of Josephson junction grounding the node, respectively. Here, the flux is expressed in units of the reduced flux quantum $\phi_0 = h/(2e)$ where $e$ is the electron charge.


Next, we introduce conjugated variables $\pi_i\equiv\dfrac{\partial\mathcal{L}}{\partial\dot{\phi_i}} = C_g \dot{\phi}_1$
and, performing a Legendre transformation $\mathcal{H}=\sum_i\pi_i\dot{\phi}_i-\mathcal{L}$, derive a Hamiltonian $\mathcal{H}(\phi,\pi)$ as
\begin{equation}
    \mathcal{H}=\frac{\pi_1^2}{2C_g} - E_g \cos{\phi_1} + \frac{(\phi_1-\phi_2)^2}{2L}+\frac{(\phi_1-\phi_3)^2}{2L}-E_J\cos{(\phi_1-\phi_4)}-E_J\cos{(\phi_1-\phi_5)}
\end{equation}
We can decompose Josephson energy in Taylor series up to first nonlinear term
\begin{equation*}
    cos(\phi_i-\phi_j)\approx 1-\frac{1}{2}(\phi_i-\phi_j)^2+\frac{1}{24}(\phi_i-\phi_j)^4
\end{equation*}
and rewrite the Hamiltonian
\begin{multline}
    \mathcal{H}=\frac{\pi_1^2}{2C_g} - E_g \left(1-\frac{\phi_1^2}{2}+\frac{\phi_1^4}{24}\right) + \frac{(\phi_1-\phi_2)^2}{2L}+\frac{(\phi_1-\phi_3)^2}{2L}-E_J\left(1-\frac{1}{2}(\phi_1-\phi_4)^2+\frac{1}{24}(\phi_1-\phi_4)^4\right)\\-E_J\left(1-\frac{1}{2}(\phi_1-\phi_5)^2+\frac{1}{24}(\phi_1-\phi_5)^4\right)
\end{multline}

Next, we introduce creation and annihilation operators $a_i,a^\dag_i$ as
\begin{align}
    \phi_i&=\sqrt{\frac{Z}{2}}(a^\dag_i+a_i)\:,\nonumber\\
    \pi_i&=i\sqrt{\frac{1}{2Z}}(a^\dag_i-a_i)\:,
    \label{impedance}
\end{align}
where the impedance $Z$ in Eq.~\eqref{impedance} is chosen such that the terms $a_i^2, (a_i^\dag)^2$ cancel out in the Hamiltonian. 

After making this substitution we drop terms that does not conserve the number of excitations (rotating wave approximation). Josephson junctions in the scheme give rise to the multiple nonlinear terms in the Hamiltonian:
\begin{multline}
    \mathcal{H}=E_0 +f_0\hat{n}_1 + J_L(\hat{a}_1^\dagger\hat{a}_2+\hat{a}_1^\dagger\hat{a}_3+\text{H.c.})
+J_J(\hat{a}_1^\dagger\hat{a}_4+\hat{a}_1^\dagger\hat{a}_5+H.c.)+\frac{E^k}{2}\hat{n}_1(\hat{n}_1-1)+\frac{T}{2}(\hat{a}_1^\dagger\hat{a}_1^\dagger\hat{a}_4\hat{a}_4+\hat{a}_1^\dagger\hat{a}_1^\dagger\hat{a}_5\hat{a}_5+H.c.)\\+\frac{J^D}{\sqrt{2}}(\hat{a}_1^\dagger(\hat{n}_1+\hat{n}_4)\hat{a}_4+\hat{a}_1^\dagger(\hat{n}_1+\hat{n}_5)\hat{a}_5+H.c.)+E^{ck}(\hat{n}_1\hat{n}_4+\hat{n}_1\hat{n}_5)
\label{eq:bulk_node}
\end{multline}
where
\begin{eqnarray}
E_0 &=& - E_g - 2E_J + (C_g Z)^{-1}/2\nonumber\\
f_0 &=& (C_g Z)^{-1}-Z^2(E_g+4E_J)/8\nonumber\\
E^k &=& -Z^2(E_g + 2 E_J)/8\nonumber\\
E^{ck} &=& -Z^2E_J/4\nonumber\\
T &=& -Z^2E_J/8\nonumber\\
J^D &=& =Z^2E_J/4\sqrt{2} \nonumber\\
J_L &=& -Z/2L\nonumber\\
J_J &=& -ZE_J/2+E_JZ^2/4 \nonumber\\
Z &=& \frac{1}{\sqrt{C_g (E_g + 2 (1/L + E_J))}}\:.
\label{eq:param_bn}
\end{eqnarray}

\subsection{The edge and corner nodes}

The equations for the edge and corner nodes of the circuit are different because of the lack of neighbors. According to the scheme of the system (Fig.\ref{fig:scheme}), edge nodes are lacking one Josephson junction while corner sites are connected only via inductances.  Following the derivation used for the bulk case, we obtained the Hamiltonian Eq.~\eqref{eq:bulk_node} with parameters modified for the edge:
\begin{eqnarray}
E_0^{(e)} &=& - E_g^{(e)} - E_J + (C_g^{(e)} Z^{(e)})^{-1}/2\nonumber\\
f_0^{(e)} &=& (C_g^{(e)} Z^{(e)})^{-1}-(Z^{(e)})^2(E_g^{(e)}+2E_J)/8\nonumber\\
E^k_{(e)} &=& -(Z^{(e)})^2(E_g^{(e)} + E_J)/8\nonumber\\
Z^{(e)} &=& \frac{1}{\sqrt{C_g^{(e)} (E_g^{(e)} + 2/L + E_J)}}\:,
\end{eqnarray}
and corner sites:
\begin{eqnarray}
E_0^{(c)} &=& - E_g^{(c)} + (C_g^{(c)} Z^{(c)})^{-1}/2\nonumber\\
f_0^{(c)} &=& (C_g^{(c)} Z^{(c)})^{-1}-(Z^{(c)})^2E_g^{(c)}/8\nonumber\\
E^k_{(c)} &=& -(Z^{(c)})^2E_g^{(c)}/8\nonumber\\
Z^{(c)} &=& \frac{1}{\sqrt{C_g^{(c)} (E_g^{(c)} + 2/L)}}\:.
\end{eqnarray}

According to Eqs.\eqref{eq:param_bn} all the couplings in the system are functions of $Z$. Thus, it seems reasonable to modify grounding elements for edge $E_g^{(e)}, C_g^{(e)}$ and corner $E_g^{(c)}, C_g^{(c)}$ sites such that $Z^{(c)}=Z^{(e)}=Z$ which ensures that the couplings for the bulk, edge and corner nodes are the same. Also we require that doublon eigenenergy is the same for all sites in the effective description. This results in the following $E_g^{(e)}, C_g^{(e)}$:
\begin{align}
    E_g^{(e)}&=E_g+\frac{16-5Z}{16-3Z}E_J\\
    C_g^{(e)}&=\frac{1}{Z(E_g^{(e)}+2/L+E_J)}
\end{align}
The same calculation for $E_g^{(c)}, C_g^{(c)}$ yields
\begin{align}
    E_g^{(c)}&=E_g+2\frac{16-5Z}{16-3Z}E_J\\
    C_g^{(c)}&=\frac{1}{Z(E_g^{(e)}+2/L)}
\end{align}

\newpage
\section{Symmetry of the eigenstates}
In this Note, we explore the symmetry of the single- and two-photon eigenmodes. We start from the two-photon case to provide a comparison and show the change in the topological invariant.  

First we revisit the properties of the rotation operator $\hat{R}_3$:
\begin{eqnarray}
\hat{R}_3 &=& \left(
\begin{array}{ccc}
     0&0&1  \\
     1&0&0  \\
     0&1&0  
\end{array}
\right)\:.
\end{eqnarray}
Its eigenvalues and the respective eigenvectors are presented in Table~\ref{tab:ro}.

\begin{table}[ht]
\centering
\begin{tabular}{|c  c  c|}
  \hline
     eigenvalue & eigenvector & p \\
     \hline
     $e^{2\pi i(p-1)/3}=1$ & $(1, 1, 1)$  & 1\\ 
     $e^{2\pi i(p-1)/3}=e^{2\pi i/3}$ & $(e^{-2\pi i/3}, e^{2\pi i/3},1)$ & 2\\
     $e^{2\pi i(p-1)/3}=e^{4\pi i/3}$ & $(e^{2\pi i/3}, e^{-2\pi i/3},1)$ & 3\\
     \hline
\end{tabular}
\caption{The eigenvalues and eigenvectors of the rotation operator $\hat{R}_3$.}
\label{tab:ro}
\end{table}

Next, we consider the reduced Bloch Hamiltonian Eq.~\eqref{eq:bloch_hamiltonian}. The doublon part of the matrix obeys $C_3$ symmetry being characterized by the $p$ symmetry index. The structure of the doublon part of the eigenmodes of the Hamiltonian in the $\Gamma$ and $K$ points of the Brillouin zone is presented in Table \ref{tab:doublon}. 

\begin{table}[ht]
    \centering
    \begin{tabular}{|c|c|}
    \hline
    $\Gamma$-point  (0,0) & K-point  ($4\pi/3$,0)\\
    \hline
\begin{tabular}{c c c}
     $f$ & $\psi$ & p \\
     \hline\\
     4.645 & $\left(e^{-2\pi i/3},  e^{2\pi i/3}, 1\right)$ & 2\\
     \\
     4.643 & $\left(e^{2\pi i/3},  e^{-2\pi i/3}, 1\right)$ & 3\\
     \\
     4.334 & $\left(1, 1, 1\right)$ & 1\\
\end{tabular}&
\begin{tabular}{c c c}
     $f$ & $\psi$ & p \\
     \hline\\
     4.645 & $\left(e^{2\pi i/3},  e^{-2\pi i/3}, 1\right)$ & 3\\
     \\
     4.643 & $\left(1, 1, 1\right)$ & 1\\
     \\
     4.335 & $\left(e^{-2\pi i/3},  e^{2\pi i/3}, 1\right)$ & 2\\
\end{tabular}\\
\hline
$\#\Gamma_1^{(3)}=1$&$\#K_1^{(3)}=0$\\
$\#\Gamma_2^{(3)}=0$&$\#K_2^{(3)}=1$\\
\hline
    \end{tabular}
    \caption{The eigenmodes of the doublon part of Bloch Hamiltonian for the non-tirvial topological regime.}
    \label{tab:doublon}
\end{table}

According to the definition of the topological invariant:
\begin{eqnarray}
\chi^{(3)} = (\#K_1^{(3)}-\#\Gamma_1^{(3)},\#K_2^{(3)}-\#\Gamma_2^{(3)})\:,
\end{eqnarray}
the doublon topological invariant equals $\chi^{(3)} = (-1,1)$ indicating non-trivial topological phase.


Then, we examine a single-photon case described by the Bloch Hamiltonian:
\begin{eqnarray}
H_{sp}&=&\left[
\begin{array}{ccc}
    f_0& J_L+J_Je^{i\phi} & J_L+J_Je^{i(\phi-\theta)} \\
    J_L+J_Je^{-i\phi} & f_0 & J_L+J_Je^{-i\theta} \\
    J_L+J_Je^{i(\theta-\phi)} & J_L+J_Je^{i\theta}  &f_0
\end{array}
\right]\:.
\end{eqnarray}
Solving the eigenvalue problem with considered parameters of the qubit network, we obtain the structure of the eigenstates in the high-symmetry points of the Brillouin zone. Their explicit form is shown in Table \ref{tab:sp} suggesting that the single-photon topological invariant equals $\chi^{(3)} = (0,0)$ which means topologically trivial system.

\begin{table}[ht]
    \centering
    \begin{tabular}{|c|c|}
    \hline
    $\Gamma$-point  (0,0) & K-point  ($4\pi/3$,0)\\
    \hline
\begin{tabular}{c c c}
     $f$ & $\psi$ & p \\
     \hline\\
     2.803 & $\left(e^{-2\pi i/3},  e^{2\pi i/3}, 1\right)$ & 2\\
     \\
     2.803 & $\left(e^{2\pi i/3},  e^{-2\pi i/3}, 1\right)$ & 3\\
     \\
     2.761 & $\left(1, 1, 1\right)$ & 1\\
\end{tabular}&
\begin{tabular}{c c c}
     $f$ & $\psi$ & p \\
     \hline\\
     2.809 & $\left(e^{-2\pi i/3},  e^{2\pi i/3}, 1\right)$ & 2\\
     \\
     2.803 & $\left(e^{2\pi i/3},  e^{-2\pi i/3}, 1\right)$ & 3\\
     \\
     2.755 & $\left(1, 1, 1\right)$ & 1\\
\end{tabular}\\
\hline
$\#\Gamma_1^{(3)}=1$&$\#K_1^{(3)}=1$\\
$\#\Gamma_2^{(3)}=0$&$\#K_2^{(3)}=0$\\
\hline
    \end{tabular}
\caption{The eigenmodes of the single-photon Bloch Hamiltonian in the trivial topological phase.}
    \label{tab:sp}
\end{table}

\end{document}